\documentclass[aps,twocolumn,prd,superscriptaddress]{revtex4-1}

\usepackage[utf8]{inputenc}

\usepackage{mathtools}
\usepackage{amsfonts}
\usepackage{mathrsfs}
\usepackage{bbm}
\usepackage{slashed}

\usepackage{graphicx}
\usepackage[dvipsnames]{xcolor}
\usepackage{array}

\usepackage{placeins}

\usepackage{xspace}
\usepackage{hyperref}

\usepackage{xifthen}
\usepackage{dsfont}
\usepackage{appendix}
\usepackage{enumitem}
\usepackage{booktabs}
\usepackage{units}
\usepackage{comment}

\setkeys{Gin}{width=0.48\textwidth}

\graphicspath{
	{./figures/}
}

\newcommand{\imag}{\text{i}}

\newcommand{\mbf}{\mathbf}

\def\0#1#2{\frac{#1}{#2}}

\def\Eq#1{Eq.~\eqref{#1}}

\newcommand{\Tr}{\mathrm{Tr}}

\def\CP{{\mathcal P}}

\def\CC{{\mathcal C}}


\newcommand{\gettitle}{Gauge-invariant condensation in the nonequilibrium quark-gluon plasma}

\hypersetup{
	colorlinks,
	linkcolor={red!75!black},
	citecolor={blue!75!black},
	urlcolor={blue!75!black},
	pdftitle={\gettitle},
	pdfauthor={Berges, Boguslavski,Pawlowski},
	pdfkeywords={Spectral function} {Non-thermal fixed points}
	{Order parameters} {Gluon} 
	{Scaling}
	{Real time} {Infrared},
	bookmarksopen=true,
	bookmarksopenlevel=2,
	bookmarksnumbered=true
}

\begin{document}

\title{\gettitle}

\author{Jürgen Berges} 
\affiliation{Institute for Theoretical Physics, 
  Heidelberg University, Philosophenweg 12, 69120 Germany}

\author{Kirill Boguslavski} \affiliation{Institute for Theoretical
  Physics, Technische Universit\"{a}t Wien, 1040 Vienna, Austria}
\affiliation{Department of Physics, University of Jyv\"{a}skyl\"{a}, P.O. Box 35, 40014, Jyv\"{a}skyl\"{a}, Finland}

\author{Mark Mace}\affiliation{Department of Physics, University of Jyv\"{a}skyl\"{a}, P.O. Box 35, 40014, Jyv\"{a}skyl\"{a}, Finland}\affiliation{Helsinki Institute of Physics, University of Helsinki, P.O. Box 64, 00014, Helsinki, Finland}

\author{Jan~M.~Pawlowski} 
\affiliation{Institute for Theoretical Physics, 
  Heidelberg University, Philosophenweg 12, 69120 Germany}

\begin{abstract}
The large density of gluons, which is present shortly after a nuclear collision at very high energies, can lead to the formation of a condensate. We identify a gauge-invariant order parameter for condensation based on elementary non-perturbative excitations of the plasma, which are described by spatial Wilson loops. Using real-time lattice simulations, we demonstrate that a self-similar transport process towards low momenta builds up a macroscopic zero mode. Our findings reveal intriguing similarities to recent discoveries of condensation phenomena out of equilibrium in table-top experiments with ultracold Bose gases.
\end{abstract}

\maketitle

%%%%%%%%%%%%%%%%%%%%%%%%%%%%%%%%%%%%%%%%%%%%%%%%%%%%%%%%%%%%%%%%%

\section{Introduction}

In high-energy collider experiments with heavy nuclei, the far-from-equilibrium matter formed shortly after the collision is expected to have a high gluon density~\cite{Gelis:2010nm,Lappi:2006fp}. 
It has been argued that this initial over-occupation of gluons may be so large that -- if the system were in thermal equilibrium with the same energy density -- a condensate would be needed to account for the excess of gluons~\cite{Blaizot:2011xf}. The possibilities for a condensate of gluon fields have been discussed in detail~\cite{Floerchinger:2013kca}. However, the
subject was disputed in view of simulation results for the plasma's evolution, 
which do not support Bose condensation of gluon fields 
~\cite{Kurkela:2012hp,York:2014wja,Blaizot:2016iir}. The analysis is complicated by the fact that the underlying theory of quantum chromodynamics (QCD) is a gauge theory. Physical
observables are gauge invariant, and examples of gauge-invariant operators for nonequilibrium condensation have been studied in the Abelian Higgs model, and its relation to non-Abelian theories~\cite{Gasenzer:2013era,Ford:1998bt,Mitreuter:1996ze}. 

In this work, we demonstrate that initial over-occupation of gluons leads to the formation of a  gauge-invariant condensate. The definition of the latter takes into account that the infrared excitations of non-Abelian gauge theories are extended objects, which can be computed from 
Wilson loops~\cite{Berges:2007re,Dumitru:2014nka,Mace:2016svc,Berges:2017igc}. Condensation is signaled by the formation of a macroscopic zero-mode expectation value, which scales proportional to \mbox{$(2\pi)^d\delta^{(d)}(0) \rightarrow L^d$} for a $d$-dimensional finite volume with length scale $L$~\cite{Berges:2012us}. Analyzing the nonequilibrium spatial Wilson loops, we identify a transport process towards low momenta building up a macroscopic zero mode. Performing simulations at various volumes, we demonstrate the scaling of the zero-mode expectation value with system size, such that a volume-independent condensate fraction is established for $L \rightarrow \infty$. We also perform a corresponding calculation of the infrared properties in thermal equilibrium with the same energy density, and show that no condensation is observed in this case.    

We compare our results for the non-Abelian plasma with the far-from-equilibrium dynamics
of Bose condensation for a scalar order-parameter field~\cite{Berges:2012us,Orioli:2015dxa,Chantesana:2018qsb} and find remarkable similarities.
The build-up of the scalar macroscopic zero mode is described in terms of a
self-similar behavior with universal scaling exponents~\cite{Orioli:2015dxa}. 
Table-top experiments with ultracold quantum gases now discovered
these universal transport processes building up a condensate starting from initial
over-occupation of bosonic excitations of trapped
atoms~\cite{Prufer:2018hto,Erne:2018gmz}. Comparing the characteristic infrared
scaling exponents of the Bose gas and the non-Abelian plasma, we observe an agreement 
within errors. This corroborates similar findings of universal scaling behavior in the perturbatively occupied regimes at higher momenta~\cite{Schlichting:2012es,Kurkela:2012hp,York:2014wja}, where the plasma's longitudinal expansion plays an important role~\cite{Berges:2013eia,Berges:2013fga,Berges:2014bba}.    

The non-perturbative time evolution is computed from classical-statistical
lattice gauge theory simulations, which provide an accurate non-perturbative description in the over-occupied regime~\cite{Aarts:2001yn,Berges:2013fga,York:2014wja,Mace:2016svc}. We consider the non-expanding system following Refs.~\cite{Berges:2007re,Mace:2016svc,Berges:2017igc}, which established for nonequilibrium spatial Wilson loops a self-similar area-law with a time-dependent spatial string tension scale, which decreases with time and sets the scale for the condensation phenomenon we find. 

The paper is organized as follows: in Sec.~\ref{sec:scales} we discuss the setup and relevant scales of the system. We introduce spatial Wilson loops in Sec.~\ref{sec:Wilson_loop} and discuss the new condensation phenomenon in Sec.~\ref{sec:condensation}. It is compared to classical thermal equilibrium in Sec.~\ref{sec:thermal} and to Bose condensation in scalar theories in Sec.~\ref{sec:scalars}. The phenomenon and its potential applications are further discussed in Sec.~\ref{sec:discussion} and we conclude with Sec.~\ref{sec:conclusions}.

\section{Dynamical separation of scales}
\label{sec:scales}

In high-energy nuclear collisions, the initially produced gluons are expected to have typical momenta of order the saturation scale $Q_s$, at time
$t \sim 1/Q_s$~\cite{Gelis:2010nm,Lappi:2006fp}, where we use natural
units with $\hbar = c = 1$. While the running gauge coupling $\alpha_s(Q_s)$ is small for large enough $Q_s$, the system is strongly
correlated because the gluon occupancies $f \sim\! 1/\alpha_s(Q_s)$ are
large. 

Here we consider the high-energy limit, because in this case the
non-perturbative quantum problem can be mapped onto a
classical-statistical lattice gauge theory, whose far-from-equilibrium
evolution can be rigorously studied using large-scale computer
simulations~\cite{Aarts:2001yn,Berges:2013fga}. 
The characteristic initial over-occupation is translated
into energy density $\sim\! Q_s^4/\alpha_s$ and fluctuations to
initialize the lattice gauge theory
evolution. In the following, dimensionful quantities will be given in suitable powers of $Q_s$.

We consider an $SU(N_c)$ gauge theory with number of colors $N_c$, and fields are discretized on a lattice with spacing $a_s$. We initialize them as a superposition of transversely polarized gluon fields $\mathcal{A}_j^a(t=0,\mbf p)= \sqrt{f(0,p)/(2p)}\sum_{\lambda} c^a_{\mbf p} \xi_j^{(\lambda)}(\mbf p) + c.c.$ for spatial momenta $\mbf p$ with $p = |\mbf p|$, and their time derivatives $E_j^a(t=0,\mbf p)= \sqrt{f(0,p)/(2p)}\sum_{\lambda} c^a_{\mbf p} \dot{\xi}_j^{(\lambda)}(\mbf p) + c.c.$. Here, $a = 1, \dots, N_c^2-1$ is the color index, the index $j = 1, 2, 3$ labels spatial components, $c.c.$ denotes complex conjugates, the $\xi_j^{(\lambda)}$ are the transverse polarization vectors, and the $c^a_{\mbf p}$ are complex Gaussian random numbers with vanishing mean and unit variance. For their real-time evolution, we solve the classical Heisenberg equations of motion in the temporal axial gauge $\mathcal{A}_0 = 0$ formulated in a gauge covariant way with $E^j_a(t,\mbf x)$ and link fields $U_j(t,\mbf x) = \exp\left(iga_s \mathcal{A}_j(t,\mbf x)\right)$, see, e.g., Refs.~\cite{Kurkela:2012hp,Berges:2013eia,Berges:2013fga} for details.

The large initial gluon occupancies are parametrized by $f(0,p) = Q_s / (4\pi\alpha_s\sqrt{p^2 + Q_s^2/10}\,)\, \theta(Q_s-p)$. 
With such highly occupied initial conditions for low momenta $p \leq Q_s$, 
the system approaches after a transient time a self-similar attractor regime, which is insensitive to the precise value of the coupling and to details of the initial conditions~\cite{Berges:2008mr,Kurkela:2011ti,Kurkela:2012hp,Berges:2013eia,Berges:2013fga,York:2014wja,Berges:2017igc,Boguslavski:2019fsb}. 

In the scaling regime the nonequilibrium plasma exhibits a hierarchy of scales, with a hard scale $\Lambda(t) \sim t^{1/7}$ dominating the system's energy density, a soft electric (Debye) screening scale $m_D(t) \sim t^{-1/7}$, and an ultrasoft magnetic scale $\sqrt{\sigma}(t) \sim t^{-\zeta/2}$ associated to the spatial string tension $\sigma$ that will be further analyzed in the following~\cite{Kurkela:2011ti,Kurkela:2012hp,Berges:2013fga,Lappi:2016ato,Mace:2016svc,Berges:2017igc,Boguslavski:2018beu}. While initially all characteristic momentum scales are of the same order $Q_s$, the self-similar evolution leads to a dynamical separation $\sqrt{\sigma}(t) \ll m_D(t) \ll \Lambda(t)$ as time proceeds. In particular, the ultrasoft scale approaches zero and in the following we will demonstrate that this sets the scale for the build-up of a coherent macroscopic state.  

The dynamics becomes non-perturbative below the magnetic scale at ultrasoft momenta, where the occupation numbers would be $f \sim\! 1/\alpha_s$. The notion of gauge-fixed particle numbers based on a distribution $f$ of gauge field modes is ill-posed in this regime. This preempts naive approaches to the phenomenon of condensation in gauge theories by counting occupancies of quasi-particle states. Out of equilibrium there is in general also no distribution with a well-defined chemical potential as employed in standard thermal equilibrium discussions of condensation. Of course, the phenomenon of condensation is not restricted to this, and can be identified from properties of correlation functions in strongly correlated systems in and out of equilibrium~\cite{PitaevskiiStringari}. We will approach this issue by studying the dynamics at long distances using gauge-invariant spatial Wilson loops.

\section{Spatial Wilson loop out of equilibrium.}  
\label{sec:Wilson_loop}

\begin{figure}[t]
\includegraphics[width=0.75\columnwidth]{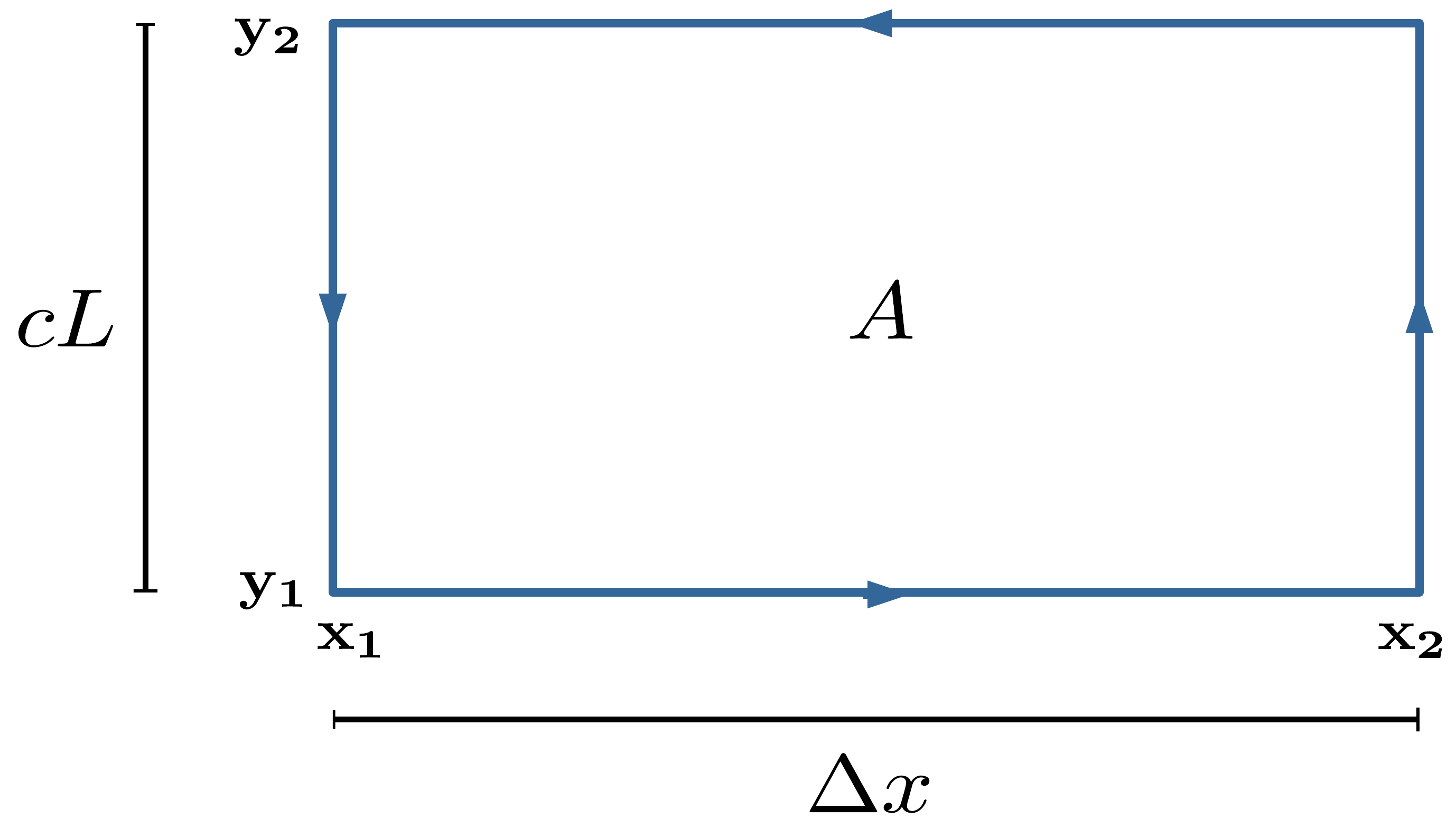}
\caption{Visualization of the rectangular path employed for 
  $\langle W(\Delta x,c L,t) \rangle$ with fixed length $\Delta y = c L$ and $c<1$. 
  }
\label{fig:Wilson}
\end{figure}

Focusing on dynamics at the magnetic scale, 
we consider the spatial Wilson loop as a gauge-invariant
quantity that captures the long-distance behavior of gauge fields
$\mathcal{A}$, defined as
\begin{align}\label{eq:Wilson}
  W(\Delta x,\Delta y,t) =\0{1}{N_c} \Tr \, \CP 
  e^{-i\, g \int_{\CC[\Delta x, \Delta y]} \mathcal{A}_i(\mbf z,t)\, dz_i} \,, 
\end{align}
where $N_c$ is the number of colors of $SU(N_c)$ gauge theory
and the index $i$ labels spatial components~\cite{Montvay:1994cy}. 
Here $\CP$ denotes path ordering, 
and the trace is in the fundamental representation. 
For simplicity, we consider rectangular paths $\CC[\Delta x, \Delta y]$
with lengths $\Delta x = |\mbf x_2 - \mbf x_1|$ and
$\Delta y= |\mbf y_2 - \mbf y_1|$ with $\mbf y_1 \equiv \mbf x_1$,
and area $A = \Delta x \Delta y$, cf.~Fig.~\ref{fig:Wilson}.

We are interested in the expectation value of the spatial Wilson
loop during the nonequilibrium evolution, which we denote by
$\langle W \rangle$. More precisely, we define our spatial Wilson loop
on an arbitrary plane on a $d=3$ dimensional cubic lattice. We
consider only on-plane Wilson loops, however, it has been
observed~\cite{Mace:2016svc} that there is no difference within available
statistics if one also includes off-plane Wilson loops. Expectation
values are obtained from averages over classical-statistical runs with
random initial seeds until convergence is observed. We also average over
fixed area loops within a single random initial seed. The Wilson loop
expectation value becomes then a function of the absolute values of
the three-dimensional vectors $\Delta \mbf x$ and
$\Delta \mbf y$.

Self-similar scaling of the far-from-equilibrium Wilson loop has been
established in Ref.~\cite{Berges:2017igc}.  Accordingly, the
nonequilibrium dynamics in this regime is described by
\begin{equation}
\label{eq:omegas}
\langle W(A,t)\rangle = \omega_S \left(A/t^{\zeta}\right)
\end{equation}
in terms of a time-independent universal scaling exponent $\zeta > 0$ and scaling function $\omega_S$. 
The positive value for $\zeta$ signals evolution towards larger length scales, with a  characteristic area $A(t) \sim t^{\zeta}$.
Moreover, for large $A/t^{\zeta}$ the scaling function obeys
\begin{equation}\label{eq:asymptoticomega}
  \lim\limits_ {(A/t^{\zeta}) \to \infty} \left(- \log \omega_S (A/t^{\zeta}) \right) 
  \sim  A/t^{\zeta} \, .
\end{equation} 
This implies a time-dependent string tension $\sigma(t) = -\partial \log \langle W\rangle / \partial A \sim t^{-\zeta}$, which can be linked to the magnetic scale and to the dynamics of topological configurations~\cite{Mace:2016svc}. The scaling exponent $\zeta$ for both $SU(2)$ and $SU(3)$ gauge theory was seen to agree at the percent level~\cite{Berges:2017igc}. 
Here we give improved estimates on $\zeta$ in \Eq{eq:zeta_from_cond} below using much later evolution times and a new approach based on the time evolution
of a zero mode that we define in the following section. Most importantly, this will allow us to study the formation of a condensate, making a link between the evolution of the spatial string tension and condensation.

\section{Gauge-invariant condensation far from equilibrium.}
\label{sec:condensation}

\begin{figure}
\includegraphics[width=1.0\columnwidth]{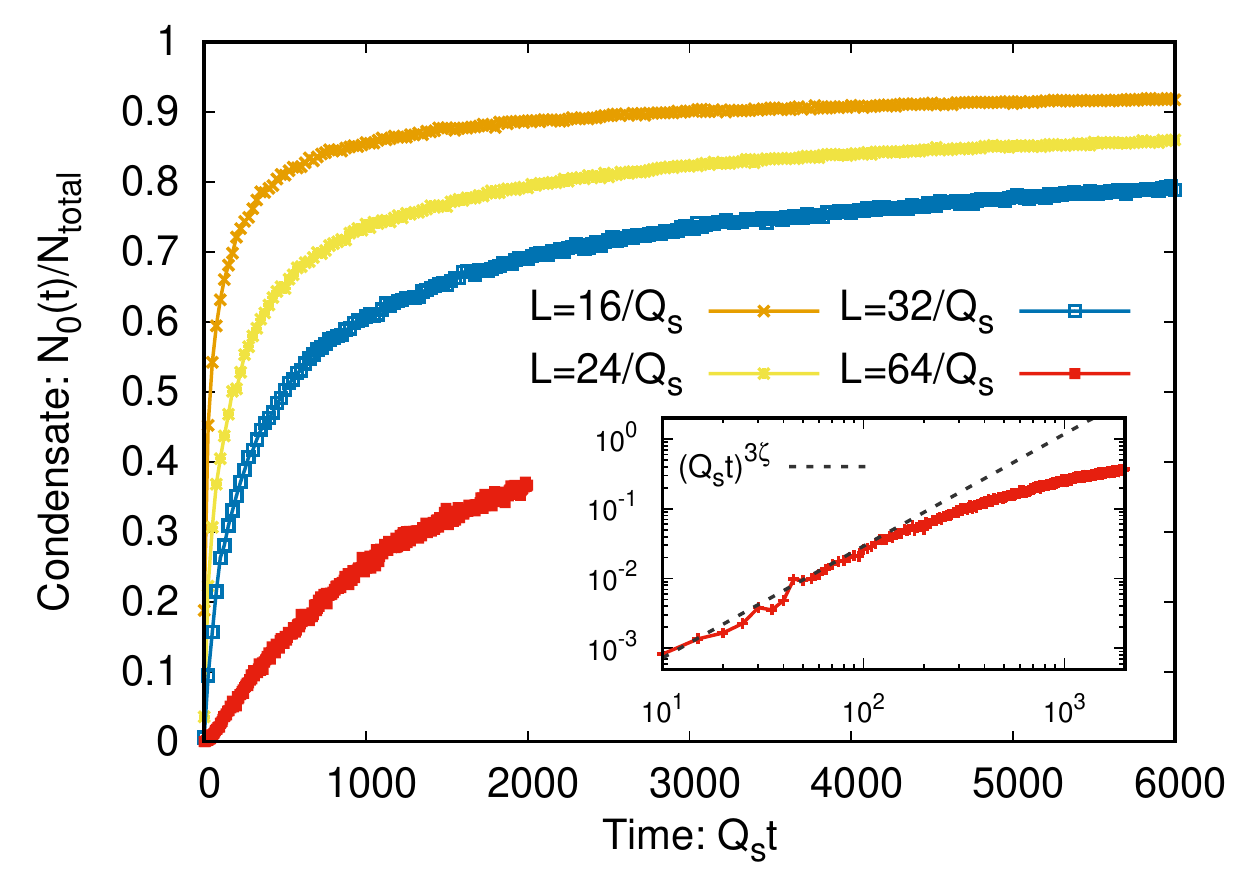}
\caption{Condensate fraction as a function of time for different volumes. Here $c=1/2$, $\zeta=0.54$, and the lattice spacing $Q_s a_s=0.5$. Inset: Logarithmic plot to demonstrate 
the early-time scaling of \Eq{eq:early}.}
\label{fig:cond}
\end{figure}

In order to study condensation, we propose to consider the closed Wilson line $\langle W(\Delta x,c L,t) \rangle$ as a function of $\Delta x$ with fixed length $\Delta y = c L$ and real parameter $c$, as illustrated in Fig.~\ref{fig:Wilson}. While $L$ denotes the entire length of the lattice, its periodicity implies that the longest physical distance for $\Delta y$ is $L/2$. Though all numerical results shown will employ $c=1/2$ accordingly, we explicitly checked that paths with $c = 1/4$ and $1/8$ lead to analogous results.

Condensation is signaled by a macroscopic 
zero mode of $\langle W(\Delta x,c L,t) \rangle$, which 
correlates the entire volume $V_c=(cL)^d$ even in the limit $L \rightarrow \infty$. We define the condensate fraction for given $L$ by integrating
with respect to $\Delta x$ and dividing through the volume $V_c$ as 
\begin{align} \label{eq:scaling_of_twopoint}
\frac{N_0(t,cL)}{N_\text{total}} \equiv \; & \frac{1}{V_c} \int_0^{cL} d^d\Delta x\, \langle W(\Delta x,cL,t)\rangle \nonumber \\
= \; &\frac{1}{V_c} \int_{0}^{cL} d^d\Delta x \,  \omega_S\! \left(\Delta x\, cL/t^{\zeta} \right) \nonumber \\
= \; & \left( \dfrac{t^{\zeta}}{(cL)^2}\right)^{\! d}\,  h\!\left((cL)^2/t^\zeta\right) \,.
\end{align}
For the second equality, we used the scaling behavior (\ref{eq:omegas}), and we define the function 
\begin{equation}
h\!\left((cL)^2/t^\zeta\right) = \int_{0}^{(cL)^2/t^{\zeta}} \!\! d^dx\,   \omega_S\! \left( x\right) \, .
\end{equation}

\begin{figure}
\includegraphics[width=1.0\columnwidth]{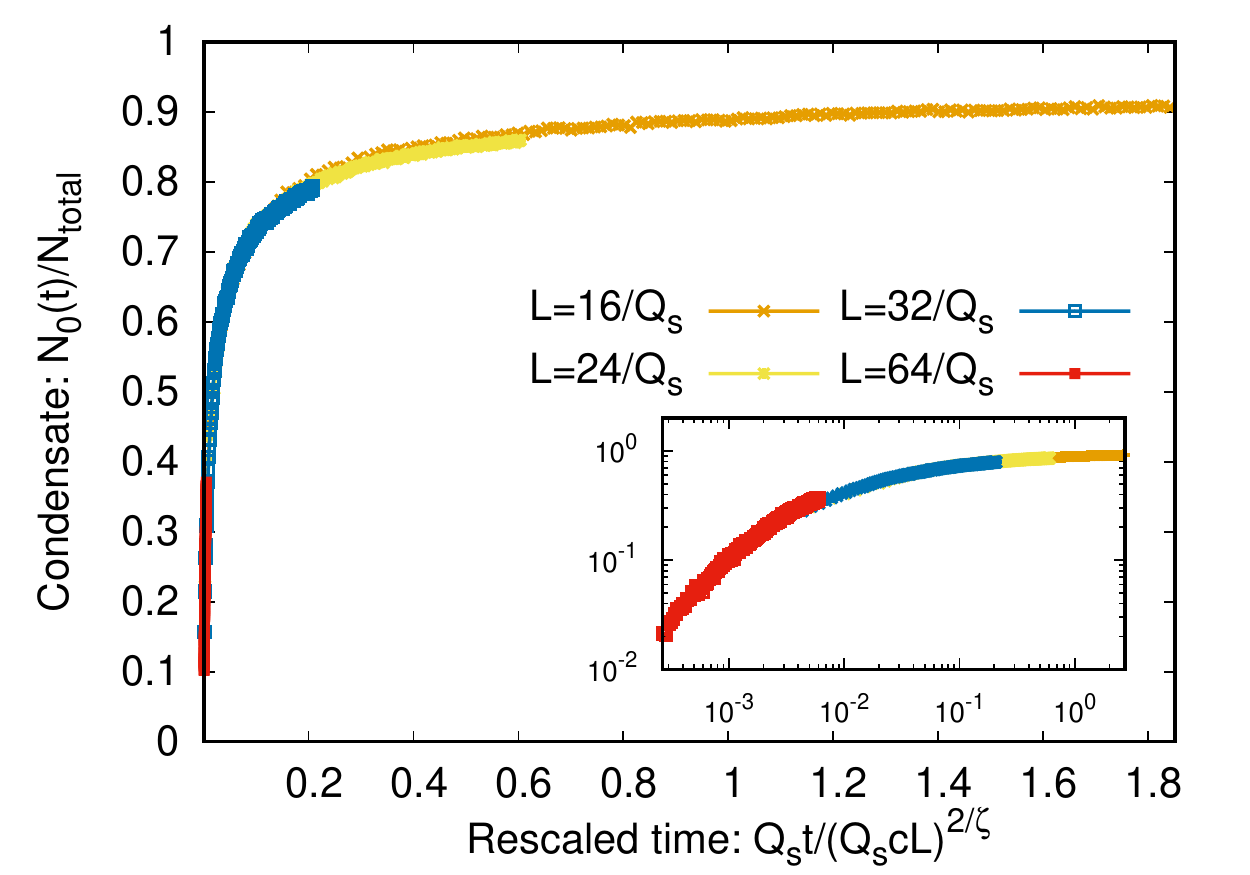}
\caption{Same as in Fig.~\ref{fig:cond} but as a function of the finite-size rescaled
  time. All curves fall on top of each other showing the emergence of a
  volume-independent condensate fraction. Inset: Same curves on a logarithmic scale.}
\label{fig:resc_cond}
\end{figure}

According to \Eq{eq:scaling_of_twopoint}, in the scaling regime the condensate fraction 
is a function of the ratio $(cL)^2/t^{\zeta}$ only. A first important case is when $(cL)^2/t^{\zeta}$ is large. This characterizes the behavior of the condensate for large enough volumes at fixed time. 
If $\omega_S$ is a rapidly decreasing function at large arguments, such as in \Eq{eq:asymptoticomega}, then
\Eq{eq:scaling_of_twopoint} takes the form
\begin{align} \label{eq:early}
\lim_{(cL)^2/t^{\zeta} \gg 1} \left( \frac{N_0(t,cL)}{N_\text{total}} \right)  \simeq \; &\left( \dfrac{t^{\zeta}}{(cL)^2} \right)^{\! d}\, h_\infty \, 
\end{align}
with the asymptotic constant
$h_\infty = \lim_{x \rightarrow \infty} h(x)$.  Since $\zeta$ is
positive, \Eq{eq:early} describes the growth of the condensate
following a power-law in time. Once the entire volume becomes
correlated, the condensate growth is expected to terminate, and is
bounded by
\begin{align}\label{eq:bound}
 \frac{N_0(t,cL)}{N_\text{total}} \leq 1 \,, 
\end{align}
since the Wilson loop satisfies $W\leq 1$.

We verify the above parametric estimates by our lattice simulation
data in $d=3$ spatial dimensions for $SU(2)$. The condensate fraction is shown in
Fig.~\ref{fig:cond} as a function of time for different volumes with
$32^3$, $48^3$, $64^3$ and $128^3$ lattice sites. The lattice spacing
is $ a_s=0.5$ in units of $Q_s$, and we have checked the insensitivity
of our results to this choice. Since we simulate at finite volumes, an initial growth of the zero mode has to cease once the entire volume of the system becomes correlated. 
Accordingly, one observes from the data that the zero mode 
grows with time, then leveling off. In the inset, the system
with the largest volume reveals the power-law growth of \Eq{eq:early}
at early times.

From the results displayed in Fig.~\ref{fig:cond} it is still not obvious that the limit $\lim_{L \rightarrow \infty} N_0(t)/N_{\text{total}}$ approaches a non-vanishing value, which is required to demonstrate condensation in the infinite volume limit. Of course, limited resources do not allow the study of infinite volumes, but the question of condensation can be settled using standard finite-size scaling analysis~\cite{Berges:2012us,Orioli:2015dxa}. 

\begin{figure}
\includegraphics[width=1.0\columnwidth]{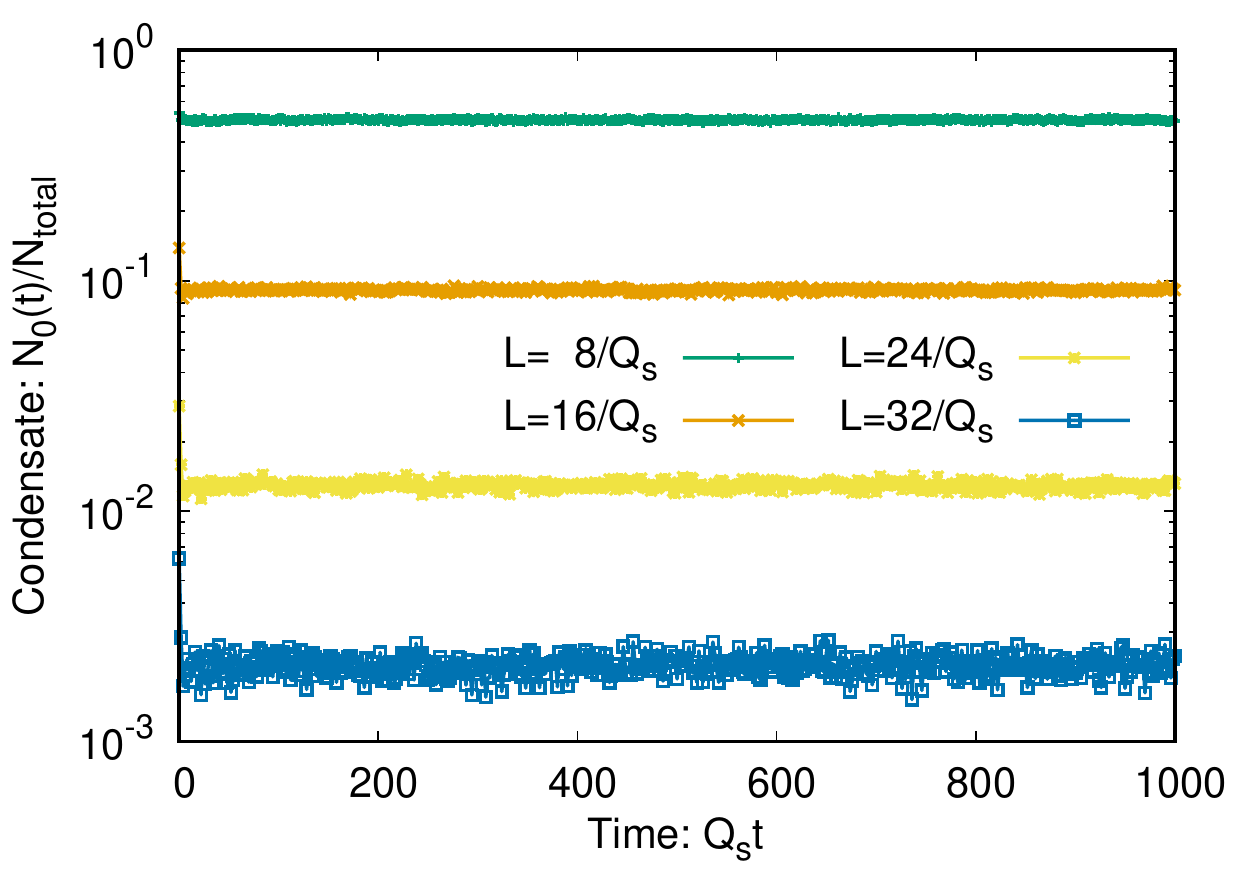}
\caption{Condensate fraction as a function of time for different volumes for systems in classical thermal equilibrium.}
\label{fig:class_therm}
\end{figure}

The scaling of the condensate formation time, $t_{\text{cond}}$, with
volume can be estimated by
\begin{align}\label{eq:gaugecondtime}
 t_{\text{cond}} \sim \left(cL\right)^{2/\zeta} \, ,
\end{align}
using the power-law growth of \Eq{eq:early}
at early times.
As a consequence, according to \Eq{eq:scaling_of_twopoint} the
evolution of the condensate fraction $N_0(t,cL)/N_\text{total}$ is
then only a function of the ratio $t/t_{\text{cond}}$. This is
demonstrated in Fig.~\ref{fig:resc_cond}, where the same curves as in
Fig.~\ref{fig:cond} are shown as functions of the rescaled time
$t/t_{\text{cond}}$ in linear and logarithmic scaling in the main
panel and in the inset, respectively. All curves fall on top of each
other, which shows the emergence of a volume-independent
condensate fraction, approaching $N_0(t,cL)/N_\text{total} \simeq 0.92$ at the latest simulation times.
Employing the $\chi^2$-procedure outlined in
Ref.~\cite{Berges:2013fga}, we extract
\begin{align}
\label{eq:zeta_from_cond}
    \zeta = 0.54 \pm 0.04\; \text{(stat.)} \pm 0.05\; \text{(sys.)}\,,
\end{align}
where systematic uncertainty results from variation of the length
fraction $c$.  This value for $\zeta$ agrees within errors with the
previously measured value using the self-similarity of Wilson
loops~\cite{Berges:2017igc}, and from extracting the string
tension~\cite{Mace:2016svc}.

\section{Classical thermal equilibrium}
\label{sec:thermal}

The classical-statistical approach cannot be used to describe the approach of the system to thermal equilibrium at late times, since the dynamics of low-occupied high-momentum modes is not properly described in this approximation~\cite{Berges:2013lsa}. Since we are interested in very low-momentum properties, it is nevertheless an insightful control case to study what happens if we directly start in or close to thermal equilibrium. In the deep infrared, the thermal Bose-Einstein statistics of the quantum theory becomes very similar to classical statistics, with $1/(\exp(\omega/T)-1) \simeq T/\omega$ for frequencies $\omega$ much below the temperature $T$ in natural units. 

Here, we compute the evolution of the same quantity \eqref{eq:scaling_of_twopoint} as before, but this time starting from classical thermal equilibrium. Results are shown in Fig.~\ref{fig:class_therm} for $c=1/8$ with lattice spacing $Q_s a_s=1$ and initial condition $f(0,p) = Q_s / p$, mimicking a state close to classical thermal equilibrium, but with different volumes with $16^3$, $32^3$, $48^3$ and $64^3$ lattice sites. Since the lattice spacing and energy density is the same for all simulations, the resulting thermal state has the same temperature. As expected, no time evolution of the quantity $N_0(t,cL) / N_\text{total}$ is visible in the figure shortly after initialization, because of time-translation invariance of thermal equilibrium.  

Most importantly, one observes that the value of $N_0(t,cL) / N_\text{total}$ decreases with volume,
approaching zero in the infinite volume limit. This is the expected behavior of a zero-mode in the absence of condensation. Of course, no finite-size rescaling of time, as performed for the over-occupied non-equilibrium case in Fig.~\ref{fig:resc_cond}, can modify this. We conclude that the condensation, which we uncovered for the over-occupied plasma, is a transient phenomenon: The build-up of the non-equilibrium condensate is a consequence of the dynamical transport process of excitations of the spatial Wilson loop towards low momenta. At the same time, there exists an energy cascade towards high momenta associated to the hard scale evolution $\Lambda(t) \sim t^{1/7}$. Thermalization is expected to set in after the time $t_*Q_s \sim 1/\alpha^{7/(d+1)}$ when the typical occupancy at this scale approaches unity~\cite{Schlichting:2012es,Kurkela:2012hp}.

\section{Comparison to Bose condensation of scalar fields}  
\label{sec:scalars}

We now compare our findings for the gauge theory to the theoretically and
experimentally established condensation dynamics in ultracold Bose
gases far from
equilibrium~\cite{Berges:2012us,Orioli:2015dxa,Prufer:2018hto}.  Here
we consider the example of an interacting Bose gas in three spatial
dimensions described by a complex scalar order-parameter field
$\phi(t,\mathbf{x})$. The s-wave scattering length $a$ and density
$n=N_\text{total}^\phi/V$ of the Bose gas can be used to define a
characteristic momentum scale $Q = \sqrt{16 \pi a n}$. In this setup,
$Q$ plays a similar role as the saturation scale for gluons in the
gauge theory case, and the diluteness $\sqrt{n a^3}$ provides the
dimensionless coupling parameter. In the dilute regime, where
$\sqrt{n a^3} \ll 1$, an over-occupied Bose gas features large
occupancies $\sim 1/\sqrt{n a^3}$ for modes with momenta of order~$Q$.

The nonequilibrium dynamics for scalars starting from over-occupation
has been studied in great detail~\cite{Berges:2008wm,Scheppach:2009wu,Berges:2010ez,%
  Nowak:2010tm,Nowak:2011sk,Berges:2012us,Berges:2014bba,%
  Orioli:2015dxa,Moore:2015adu,Walz:2017ffj,Chantesana:2018qsb,Deng:2018xsk}. For
spatially translation invariant systems, the infrared regime exhibits
the self-similar scaling behavior
\begin{equation}\label{eq:phiphi}
  \frac{\langle \left\lbrace \phi(t,\mathbf{x}_1),\phi^\dagger(t,\mathbf{x}_2) \right\rbrace \rangle}{\langle \left\lbrace \phi(t,0), \phi^\dagger(t,0) \right\rbrace\rangle}  =
  f_S(\Delta x/t^{\beta}).
\end{equation}
Here $\langle \left\lbrace \phi,\phi^\dagger \right\rbrace \rangle$ is the connected part of the anticommutator correlation and $f_S$ denotes the scaling function, with scaling exponent \cite{Orioli:2015dxa,Schachner:2016frd,Chantesana:2018qsb}
\begin{equation}\label{eq:beta}
\beta = 0.55 \pm 0.05 \, .
\end{equation}   
The positive value for $\beta$ signals evolution towards larger
scales, with characteristic length $\Delta x(t) \sim
t^{\beta}$. Accordingly, this corresponds to scaling towards low
momentum modes in Fourier space. Asymptotically, the
scaling function obeys~\cite{Mikheev:2018adp}
\begin{equation}
\lim\limits_{\left(\Delta x/t^{\beta}\right) \to \infty} \left(- \log f_S(\Delta x/t^{\beta}) \right) 
\sim  \Delta x/t^{\beta} \, .
\end{equation}

Using that
$N_\text{total}^\phi = \int_{0}^{L} d^dx\,\langle \left\lbrace\phi(t,\mbf x), \phi^\dagger(t,\mbf x) \right\rbrace \rangle / 2
= V \langle \left\lbrace\phi(t,0), \phi^\dagger(t,0) \right\rbrace \rangle / 2$ 
is conserved in the non-relativistic system, the condensate
fraction is
\begin{align} \label{eq:scalarscaling}
\frac{N_0^\phi(t)}{N_\text{total}^\phi} 
= \; &\frac{1}{V} \int_{0}^{L} d^dx\, \frac{\langle \left\lbrace\phi (t,x), \phi^\dagger(t,0) \right\rbrace\rangle}{\langle \left\lbrace\phi(t,0), \phi^\dagger(t,0) \right\rbrace \rangle} \nonumber \\
= \; &\frac{1}{V} \int_{0}^{L} d^d\Delta x \,  f_S\! \left(\Delta x/t^{\beta} \right) \nonumber \\
= \; & \left( \dfrac{t^{\beta}}{L}\right)^{\! d}\,  h^\phi\!\left(L/t^\beta\right) \,,
\end{align}
with  
\begin{equation}
h^\phi\!\left(L/t^\beta\right) = \int_{0}^{L/t^{\beta}} \!\! d^dx\,   f_S\! \left( x\right) \, .
\end{equation}
We note that the condensate fraction indeed satisfies $N_0^\phi(t) / N_\text{total}^\phi \leq 1$.
Following along the lines of the discussion for the gauge theory, the condensation time for scalars then scales as
\begin{align} \label{eq:scalarcondtime}
 t_{\text{cond}} \sim L^{1/\beta} \, ,
\end{align}
and the scalar system exhibits an early-time power law growth of the condensate fraction with $t^{\beta d}/L^d$ for large volumes, subsequently approaching a finite value~\cite{Orioli:2015dxa}. 

One observes that practically all of the above equations for scalars have precise corresponding expressions in the gauge theory, such as (\ref{eq:scalarscaling}) replacing (\ref{eq:scaling_of_twopoint}). Comparing these equations, there is an apparent difference concerning the $L^2$-dependence of the gauge theory expressions, whereas the corresponding ones for scalars depend on $L$. This additional power of $L$ appears because we chose for the gauge theory $\Delta y$ to scale with $L$.   
Instead, we could also assign a fixed extent to $\Delta y$. This does not change the condensation phenomenon we are reporting here, but will merely change the scaling with $L$. For instance, for fixed $\Delta y$ the condensate formation time scales with the length as $t_{\text{cond}}^{\Delta y = \text{const}} \sim \left(cL\right)^{1/\zeta}$ in complete analogy to (\ref{eq:scalarcondtime}), which we confirmed numerically for $\Delta y = 8$. Moreover, in both theories there is a corresponding conserved quantity. In the scalar case it is given by the conserved particle number density $\sim \!\langle \left\lbrace\phi(t,0), \phi^\dagger(t,0) \right\rbrace \rangle$, while in the gauge theory this role is played by $\langle W(\Delta x = 0,cL,t)\rangle = \text{const}$.

\section{Discussion}
\label{sec:discussion}

In view of this close correspondence, it is remarkable that even the values for the infrared scaling exponents $\zeta$ in \Eq{eq:zeta_from_cond} and $\beta$ in \Eq{eq:beta} agree well within errors. This is highly non-trivial, since we are comparing relativistic and non-relativistic systems with different symmetry groups and field content. However, though we have considered the example of a non-relativistic Bose gas, the same infrared scaling and condensation properties have been established for relativistic $N$-component real scalar field theories~\cite{Orioli:2015dxa}. Even the anisotropic dynamics of relativistic scalars with longitudinal expansion along the $z$-direction, relevant in the context of heavy-ion collision kinematics, shows a very similar condensation behavior~\cite{Berges:2015ixa}. Because of the strong  enhancement in the over-occupied infrared regime, the low momentum modes exhibit essentially isotropic properties. We therefore expect longitudinally expanding non-Abelian plasmas, commonly employed in early-time descriptions of relativistic heavy-ion collision, to also exhibit the condensation phenomenon reported here.

In view of applications to the construction of effective descriptions such as hydrodynamics, it is advantageous to link the condensation phenomenon observed here in terms of the traced Wilson loop directly to correlation functions of a gauge invariant scalar field. 
In Ref.~\cite{Gasenzer:2013era} it has been shown how to link a non-Abelian gauge theory to the Abelian Higgs model, where the adjoint Higgs field $\varphi$ is the algebra element of a closed spatial Wilson line. Translated to the current setup this gives $\exp (\imag\varphi(t,x_1,x_2)) = {\cal P} \exp\left[ig\int_0^L d x_3 \mathcal{A}_3(t,{\bf x})\right]$. The scalar field $\varphi$ is 
closely related to the gauge field itself as can be seen in specific gauges, such as in the Polyakov gauge employed in Refs.~\cite{Ford:1998bt,Mitreuter:1996ze}. Indeed, in equilibrium it (or rather its gauge invariant eigenvalues) serves as an (gauge invariant) order parameter for the confinement-deconfinement phase transition as shown in Refs.~\cite{Braun:2007bx,Fister:2013bh}. At sufficiently large distances the closed spatial Wilson line $W(\Delta x, c L, t)$ considered in the present work is closely related to the correlator $\langle \textrm{tr} \,\exp(\imag \varphi(t, x_1, x_2))\; \textrm{tr} \exp (-\imag \varphi(t, x_1+\Delta x, x_2))\rangle / N_c^2$, and it has been this relation that triggered the present numerical work. The proportionality factors of this relation require a careful discussion of ultraviolet divergences and renormalization, as described, e.g., in Refs.~\cite{Bazavov:2016uvm,Bazavov:2018wmo}.

\section{Conclusions}
\label{sec:conclusions}

We have demonstrated that in initially over-occupied non-Abelian gauge theory at very high energies, a macroscopic zero mode for the gauge-invariant closed spatial Wilson line emerges.
The condensate growth follows a power law at early times $\sim (t/t_{\text{cond}})^{\zeta d}$ for large volumes, which terminates when the entire volume becomes correlated. The condensate formation time $t_{\text{cond}} \sim (cL)^{2/\zeta}$ grows with system size. The scaling exponent $\zeta$ is universal, such that its value is independent of the details of the underlying microscopic parameters like coupling strength or initial conditions. The emergence of a condensate may have consequences for effective kinetic or hydrodynamic descriptions in the context of heavy-ion collisions.

Our comparison to theoretically and experimentally established condensation dynamics in scalar field theories uncovers an intriguing similarity in the infrared scaling behavior of non-Abelian gauge theory and $\text{(non-)}$relativistic scalars. 
Even the values for the universal scaling exponents agree within errors. This hints at a connection between these theories, and possible relations using, e.g., spatial Polyakov lines or low-energy effective theories can be studied.

In all these different theories, condensation arises 
in initially over-occupied systems
as a consequence of a self-similar transport process towards large distances in the presence of a conserved quantity. These robust ingredients can be found in a wide range of nonequilibrium systems from early-universe cosmology \cite{Berges:2008wm} to cold quantum gases \cite{Prufer:2018hto,Erne:2018gmz}. 

%{\it Acknowledgements.---}

\begin{acknowledgements}
We thank S.~Floerchinger, O.~Garcia-Montero, A.~Ipp, A.~Kurkela, T.~Lappi, A.~Mazeliauskas, J.~Peuron, A.~Pi\~{n}eiro Orioli, A.~Rebhan, K.~Reygers, S.~Schlichting, R.~Venugopalan for discussions and/or collaborations on related work. The work is supported by EMMI, the BMBF grant
  05P18VHFCA, and is part of and supported by the
  DFG Collaborative Research Centre SFB 1225 (ISOQUANT) as well as by
  the DFG under Germany's Excellence Strategy EXC - 2181/1 - 390900948
  (the Heidelberg Excellence Cluster STRUCTURES).
M.M.~is supported by the European Research
Council, grant ERC-2015-CoG-681707.
This research used resources of the National Energy Research Scientific Computing Center (NERSC), a U.S.\ Department of Energy Office of Science User Facility operated under Contract No.\ DE-AC02-05CH11231.
\end{acknowledgements}

\bibliography{NoneqOrderParamBib}

\end{document}